\documentclass[twocolumn,secnumarabic,amssymb, nobibnotes, aps, prd]{revtex4}

\usepackage{amsmath,amssymb}
\usepackage{graphicx}
\usepackage{setspace}
\usepackage[sort&compress]{natbib}
\usepackage{hyperref}
\hypersetup{colorlinks=true, citecolor=blue}

\begin{document}

\title{Modified $f(R,T)$ theory in light of gravitational wave standard sirens}

\author{Mahnaz Asghari}
\email{mahnaz.asghari@shirazu.ac.ir}
\affiliation{Department of Physics, College of Science, 
	Shiraz University, Shiraz 71454, Iran \\
	Biruni Observatory, College of Science, 
	Shiraz University, Shiraz 71454, Iran} 
\author{Ahmad Sheykhi}
\email{asheykhi@shirazu.ac.ir}
\affiliation{Department of Physics, College of Science, 
	Shiraz University, Shiraz 71454, Iran \\
	Biruni Observatory, College of Science, 
	Shiraz University, Shiraz 71454, Iran}

\begin{abstract}
    In this paper, we ponder observational constraints
    on the modified $f(R,T)$ gravity, where the gravitational action is
    a function of Ricci scalar $R$ plus the trace of the energy-momentum
    tensor $T$, regarding the functional form 
    $f(R,T)=R+2f(T)$ with $f(T)=8\pi G\lambda T$. 
    For this purpose, we utilize recently available data,
    including cosmic microwave background, weak lensing, supernovae,
    baryon acoustic oscillations, and redshift-space distortions
    measurements, together with forcasted gravitational wave 
    (GW) data from Laser Interferometer Space Antenna (LISA). 
    Notably, we examine the potentiality of simulated GW data 
    from LISA standard sirens (SS) sources to enhance 
    cosmological constraints on the $f(R,T)$ model parameters. 
    In this regard, we create three LISA mock catalogs, 
    namely Pop III, Delay, and No Delay, to improve the 
    obtained constraints on cosmological parameters of $f(R,T)$ 
    gravity from current observations. Numerical analysis
    reveals that mock GW data from LISA SS sources 
    make marginal improvements on constraining the $f(R,T)$ 
    model cosmological parameters. 
    
\end{abstract}

\maketitle

\section{Introduction}
The gravitational wave (GW) physics proves to be beneficial in
understanding the nature of gravity, as well as investigating
beyond the general theory of relativity (GR). Crucially, the
detection of GW event GW150914 by the
LIGO\footnote{Laser Interferometer Gravitational-Wave Observatory}
collaboration, initiated new opportunities to explore fundamental
physics \cite{ligo}. Moreover, the opening of multi-messenger
cosmology era by the LIGO-Virgo detection of the binary neutron
star merger GW170817 \cite{gw1} associated with its electromagnetic
(EM) counterpart GRB170817A \cite{gw2}, proposed a new way to
measure cosmological parameters \cite{multi1}. Actually, the
GW signal from coalescing compact binary systems can be utilized
for a direct measurement of their luminosity distance, which
makes them as standard sirens (SS), the gravitational analogue of
standard candles \cite{ss1,ss2,ss3}. In this direction, SS are
considered as effectual probes for estimating cosmological parameters
such as the Hubble constant \cite{schutz1,schutz2}. Accordingly,
the first measurement of $H_0$ based on SS analysis led to the
result $H_0=70.0^{+12.0}_{-8.0}$ $\mathrm{km\,s^{-1}\,Mpc^{-1}}$,
being independent of the electromagnetic distance scale \cite{ssobs1}.
Thereafter, the first joint GW determination of Hubble constant,
using GW170817 with its EM counterpart in conjunction with binary
black hole detections, resulted in an estimate of
$H_0=68.7^{+17.0}_{-7.8}$ $\mathrm{km\,s^{-1}\,Mpc^{-1}}$ \cite{ssobs2}.
However, large errors reported in $H_0$ measurements from SS approach,
requests more precise estimations based on next generation detectors
such as the space-borne interferometer
LISA\footnote{Laser Interferometer Space Antenna} \cite{lisa1,lisa2}.

The LISA mission is planned to study the gravitational universe in
the millihertz band (from below $10^{-4}$ Hz to above $10^{-1}$ Hz),
providing opportunities to investigate the history of the universe
prior to the epoch of cosmic reionization. Probing GWs in low-frequency
regime which is rich with massive binary mergers, enables us to test
gravity with unprecedented precision \cite{lisa1}. Therefore, it is
possible to utilize mock GW data from LISA SS sources 
as a complementary probe in pursuit of exploring alternative 
models of gravity. Cosmological studies concerning 
simulated data from LISA SS are widely discussed in 
the literature \cite{ssapp1,ssapp2,ssapp3,ssapp4,ssapp5,
	ssapp6,ssapp7,ssapp8,ssapp9}.

Certainly, in compliance with precise cosmological observations
such as type Ia supernovae (SNeIa) \cite{sn1,sn2} and cosmic microwave
background (CMB) anisotropies \cite{cmb1,cmb2,cmb3}, the standard
$\Lambda$CDM model based on GR is well known as an extremely
successful paradigm to describe the universe. Nevertheless,
the concordance cosmological model has some insufficiencies
from an observational viewpoint. In particular, the value of
Hubble constant $H_0$ deduced from local observational measurements
is in notable tension with CMB data \cite{H01,H02,H03,H04,H05,H06}.
In addition, direct estimations of the structure growth parameter
$\sigma_8$ report inconsistencies with Planck measurements
\cite{s81,s82,s83,s84,s85,s86}. Thus, the observed discrepancies
between the early and late-time determinations of cosmological
parameters, provide prospects for new physics beyond the standard
model of cosmology. In this respect, some modifications on GR are
proposed to achieve a more exhaustive theory of gravity.

In the framework of modified gravity theories, one can simply
replace the standard Einstein-Hilbert action with a more general
function of the scalar curvature $R$, introducing the $f(R)$
theory of gravity \cite{fr1,fr2}.
Furthermore, as a general extension of $f(R)$ model, it is viable
to contemplate a non-minimal matter-geometry coupling in
gravitational Lagrangian, known as $f(R,L_m)$ theory
\cite{frlm1,frlm2}. Accordingly, the interaction between curvature
and matter yields an extra force which impels massive particles
to have non-geodesic motion \cite{frlm3,frlm4}.
Thereupon, the non-minimal coupling between matter and geometry
motivated Harko et al. to propose a more general extension of GR
theory, namely $f(R,T)$ gravity \cite{frtHarko}
(where $T$ is the trace of the energy-momentum tensor).
Similar to the case of $f(R,L_m)$ theory, the massive test particles
do not follow geodesic paths in $f(R,T)$ model, and hence,
a supplementary acceleration emerges due to matter-curvature interaction
\cite{frtHarko,thermo1,thermo2}. Several investigations are
conducted with regard to $f(R,T)$ gravity.
For instance, the reconstruction of cosmological models in
$f(R,T)$ theory was studied in \cite{frtJamil}. \cite{frtAlvarenga}
explored dynamics of scalar perturbations in $f(R,T)$ model.
\cite{frtXu} studied the quantum cosmology of $f(R,T)$ gravity.
The existence of Noether symmetry in $f(R,T)$ theory was
considered in \cite{frtSharif}. \cite{frtBarrientos} discussed
the metric-affine approach to $f(R,T)$ gravity.
Thick braneworld systems in the framework of $f(R,T)$ cosmology
was also investigated in \cite{frtRaso}.
Further studies on $f(R,T)$ modified gravity can be carried out in
\cite{frt1,frt2,frt3,frt4,frt5,frt6,frt7,frt8,frt9,frt10,frt11,frt12,
    frt13,frt14}.
It is also worth turning our attention to GW investigations in the
context of $f(R,T)$ theory. In this regard, Alves et al. 
\cite{frtgw1} studied the physical features of GWs in the context of
$f(R,T)$ theory. Sharif and Siddiqa explored the propagation 
of axial GWs \cite{frtgw2} and also the propagation of polar GWs 
\cite{frtgw3} in a flat 
FLRW\footnote{Friedmann-Lema\^itre-Robertson-Walker} universe 
considering $f(R,T)$ model. Echoes of GWs from the surface of 
compact stars were investigated in $f(R,T)$ gravity by 
Bora and Goswami \cite{frtgw4}. In addition, Azizi et al. \cite{frtgw5} 
studied the propagation of GWs in a cosmological background 
through the cosmic fluid in $f(R,T)$ theory.
Moreover, it should be noted that applying solar system data
via the parameterized post-Newtonian formalism, results in 
severe constraints on $f(R,T)$ model \cite{frtss}, 
possibly because of the screening mechanism at 
solar system scales which is needed to ensure consistency 
with local gravitational tests.
Nevertheless, this result can not rule out $f(R,T)$ modified 
gravity models on cosmological scales.  
So, it is important to constrain $f(R,T)$
model with cosmological data which is considered significantly 
in the literature \cite{frtobs1,frtobs2,frtobs3,frtobs4,frtobs5,
    frtobs6,frtobs7,frtobs8},
exploiting current observational measurements.
Accordingly, contemplating SS as a powerful probe to scrutinize
modified gravity, in the present investigation we intend to make
use of forecasted GW data from LISA SS sources 
to obtain reliable constraints on the $f(R,T)$ model parameters.
Specifically, we focus on exploring the capability of 
simulated GW data from LISA SS sources to enhance existing 
observational constraints on modified $f(R,T)$ theory.  
In particular, we improve constraints on the studied
$f(R,T)$ gravity in Ref. \cite{ASH} by employing 
generated mock LISA SS catalogs along with 
current observations, namely CMB, weak lensing, supernovae,
baryon acoustic oscillations (BAO), and redshift-space
distortions (RSD) data.

The paper is structured as follows. We explain the modified field
equations based on $f(R,T)$ gravity formalism in section
\ref{sec2}. We introduce the exploited observational tests for
constraining $f(R,T)$ model, including 
current observations and simulated GW data, 
as well as discussing obtained constraints on the model parameters 
in section \ref{sec3}. The last section is devoted to closing
remarks.
%%%%%%%%%%%%%%%%%%%%%%%%%%%%%%%%%%%%%%%%%%%%%%%%%%%%%%%%%%%%%%%%%%
\section{Field equations in modified $f(R,T)$ gravity} \label{sec2}
In this part we consider modified field equations in the framework
of $f(R,T)$ theory, which is described by the following action \cite{frtHarko}
\begin{align} \label{eq1}
S=\frac{1}{16\pi G}\int{\mathrm{d}^{4}x\,\sqrt{-g}f(R,T)}+\int{\mathrm{d}^{4}x\,\sqrt{-g} L_m} \,.
\end{align}
Concerning action (\ref{eq1}), the gravitational Lagrangian is a
function of the Ricci scalar $R$ and the trace of
the energy-momentum tensor $T$ ($T=g^{\mu \nu}T_{\mu \nu}$), and
$L_m$ denotes the matter Lagrangian density. Then, modified field
equations in $f(R,T)$ model can be written as \cite{frtHarko}
\begin{align} \label{eq2}
&R_{\mu \nu}\frac{\partial f}{\partial R}-\nabla_{\mu} \nabla_{\nu}\frac{\partial f}{\partial R}
+g_{\mu \nu}\Box\frac{\partial f}{\partial R}-\frac{1}{2}fg_{\mu \nu} \nonumber \\
&=8\pi G T_{\mu \nu}-\frac{\partial f}{\partial T}\big(T_{\mu \nu}+\Theta_{\mu \nu}\big) \,,
\end{align}
where the energy content of the universe is assumed as a prefect fluid
with $T_{\mu\nu}=\big(\rho+p\big)u_{\mu}u_{\nu}+pg_{\mu \nu}$,
and $\Theta_{\mu \nu}$ defined as \cite{frtHarko}
\begin{align} \label{eq3}
\Theta_{\mu \nu}&\equiv g^{\alpha\beta}\frac{\delta T_{\alpha\beta}}{\delta g^{\mu \nu}} \nonumber \\
&=-2T_{\mu \nu}+g_{\mu \nu}L_m-2g^{\alpha\beta}\frac{\partial^2 L_m}{\partial g^{\mu \nu}\partial g^{\alpha\beta}} \,.
\end{align}
Notably, in case of a perfect fluid, the on-shell matter Lagrangian
takes at least three forms, mainly $L_m=p$, $L_m=-\rho$,
or $L_m=T$ \cite{L1}. The case $L_m=T$ is relevant to fluids with
equation of state parameter $0\leq w \leq 1/3$ and so is not
applicable to describe dark energy. Conversely, considering
$L_m=p$ and $L_m=-\rho$ is an appropriate choice for dark energy
fluid \cite{L1}. Hence, we contemplate $L_m=-\rho$ in our investigation,
where a linear dependency of the matter Lagrangian on the metric is
supposed \cite{L2}. Then, $\Theta_{\mu \nu}$ take the form
\begin{equation} \label{eq4}
\Theta_{\mu \nu}=-2T_{\mu\nu}-\rho g_{\mu \nu} \,.
\end{equation}
Furthermore, we are interested in a simple functional form of
$f(R,T)$ given by \cite{frtHarko}
\begin{equation} \label{eq5}
f(R,T)=R+2f(T) \,,
\end{equation}
with $f(T)$ defined as
\begin{equation} \label{eq6}
f(T)=8\pi G\lambda T \,,
\end{equation}
where $\lambda$ is a dimensionless constant.
Consequently, field equations in $f(R,T)$ theory can be find as
\begin{align} \label{eq7}
R_{\mu \nu}-\frac{1}{2}Rg_{\mu \nu}=8\pi G\Big((1+2\lambda)T_{\mu \nu}
+\lambda T g_{\mu \nu}+2\lambda \rho g_{\mu \nu}\Big) \,.
\end{align}
Concerning modified field equations (\ref{eq7}), this $f(R,T)$ model
resembles a gravitational model with an effective cosmological constant
\cite{frtHarko,thermo1}. In particular, the coefficient of metric on
the right hand side of (\ref{eq7}) may be counted as a time dependent
cosmological constant \cite{frt4}.
It should be noted that there is some debate on the functional 
form (\ref{eq5}) of $f(R,T)$ models (introduced by Harko et al. 
\cite{frtHarko}) in the literature, raised by Fisher \& Carlson 
\cite{fcd1,fcd3}, and Harko \& Moraes \cite{Hd2}. According to 
Fisher \& Carlson \cite{fcd1}, choosing the functional form 
$f(R,T)=f_1(R)+f_2(T)$ in modified $f(R,T)$ theory would not 
yields new physics, since the term $f_2(T)$ has no physical 
significance and should be incorporated into the matter 
Lagrangian. On the other hand, Harko \& Moraes reexamined the 
Fisher \& Carlson approach, recognizing some conceptual 
problems relevant to physical interpretation of the 
$T$-dependence in $f(R,T)$ model in Ref. \cite{fcd1}, 
and then represented a more clarified explanation on 
the functional form $f(R,T)=f_1(R)+f_2(T)$ \cite{Hd2}. 
Considering debates on the form of $f(R,T)$ in the literature, 
Panda et al. \cite{pds1,pds2} propounded a possible resolution 
within the context of K-essence geometry.

Regarding the usual $f(R,T)$ gravity, 
it is known that the energy-momentum tensor is given by 
\begin{equation} \label{eqk3}
{T}_{\mu \nu}={g}_{\mu \nu}L_m-2\frac{\partial L_m}{\partial {g}^{\mu \nu}} \,,
\end{equation}
and then choosing for example $p=L_m$ \cite{frtHarko} 
(with $p$ the pressure), results in a zero value for the 
second term of the energy-momentum tensor described in 
equation (\ref{eqk3}). Thus, it can be easily understood 
that the trace of the energy-momentum tensor 
(and correspondingly the term $f_2(T)$) is a function 
of $L_m$ ($T=g^{\mu \nu}T_{\mu \nu}=4L_m$). So, 
as discussed in \cite{Hd2,fcd1}, both $f_2(T)$ and $L_m$ 
are functions of the same arguments and one can consider 
an effective Lagrangian $L_m^{eff}=f_2(T)+L_m$ \cite{Hd2}, 
and then the action can be written as \cite{Hd2}
\begin{align} \label{eqk4}
S&=\int{\mathrm{d}^{4}x\,\sqrt{-g}\Big(\frac{1}{16\pi G}f_1(R)+f_2(T)+L_m\Big)}  \nonumber \\ 
&=\int{\mathrm{d}^{4}x\,\sqrt{-g}\Big(\frac{1}{16\pi G}f_1(R)+L_m^{eff}\Big)} \,.
\end{align}
On the other hand, in the K-essence $f(R,T)$ gravity, 
the emergent energy-momentum tensor is defined as \cite{pds1}
\begin{equation} \label{eqk1}
{T}_{\mu \nu}=\mathcal{G}_{\mu \nu}L(X)-2\frac{\partial L(X)}{\partial \mathcal{G}^{\mu \nu}} \,,
\end{equation}
where $\mathcal{G}_{\mu \nu}$ is the K-essence emergent gravity 
metric (refer to \cite{emk1,pds1} for more details), and the 
Lagrangian $L(X)$ is a function of the canonical kinetic 
term $X$ given by \cite{pds1}
\begin{equation} \label{eqk2}
X=\frac{1}{2}g^{\mu \nu}\nabla_{\mu}\phi\nabla_{\nu}\phi \,,
\end{equation}
with the K-essence scalar field $\phi$.
In this approach, when we consider for example $p=L(X)$ 
which is known as the purely kinetic K-essence model 
\cite{pkk}, then according to (\ref{eqk2}) we notice that 
the pressure $p$ depends on the gravitational metric 
$g^{\mu \nu}$ and the first derivative of $\phi$ 
(detailed discussions can be find in \cite{pds1,pds2}). 
Consequently, the second term of the ${T}_{\mu \nu}$ 
relation (\ref{eqk1}) would not be zero, and so the term 
$f_2(T)$ can not be absorbed by the Lagrangian $L(X)$.
Accordingly, we perceive that the additive form (\ref{eq5}) 
of $f(R,T)$ models is not questionable in the context of 
K-essence $f(R,T)$ gravity.

We consider a perturbed spatially flat FLRW 
metric in the synchronous gauge, where for the scalar 
mode we have
\begin{equation} \label{eq8}
\mathrm{d}s^2=a^2(\tau)\Big(-\mathrm{d}\tau^2+\big(\delta_{ij}+h_{ij}\big)\mathrm{d}x^i\mathrm{d}x^j\Big) \,,
\end{equation}
in which
\begin{align} \label{eq9}
h_{ij}(\vec{x},\tau)=\int \mathrm{d}^3k\,e^{i\vec{k}.\vec{x}}
\bigg(\hat{k}_i\hat{k}_jh(\vec{k},\tau)
+\Big(\hat{k}_i\hat{k}_j-\frac{1}{3}\delta_{ij}\Big)6\eta(\vec{k},\tau)\bigg) \,,
\end{align}
with scalar perturbations $h$ and $\eta$,
and $\vec{k}=k\hat{k}$ \cite{pt}, where $\vec{k}$ is 
the wavevector and $k$ is the wavenumber of the perturbations 
in Fourier space.
Thereupon, modified Friedmann equations at background level 
take the form
\begin{align}
& H^2=\frac{8\pi G}{3}\Big((1+\lambda)\sum_{i}\bar{\rho}_i-3\lambda \sum_{i}\bar{p}_i\Big) \,, \label{eq10} \\
& 2\frac{H'}{a}+3H^2=-8\pi G\Big(\lambda\sum_{i}\bar{\rho}_i+(1+5\lambda)\sum_{i}\bar{p}_i\Big) \,, \label{eq11}
\end{align}
where the prime indicates derivative with respect 
to the conformal time, the bar represents a quantity evaluated at 
background level, and index \textit{i}
indicates the component \textit{i}th in the universe filled with
radiation (R), baryons (B), dark matter (DM) and cosmological
constant ($\Lambda$). It is important to note that, in a universe
with the matter content is considered as dust, the first
modified Friedmann equation becomes
\begin{equation} \label{eq12}
H^2=\frac{8\pi G}{3}(1+\lambda)\sum_{i}\bar{\rho}_i \,,
\end{equation}
and consequently, we realize that in the studied $f(R,T)$ 
model, the term $2f(T)$ in the gravitational action modifies 
the gravitational interaction between matter and curvature, 
where accordingly the gravitational constant $G$ is 
replaced by a running gravitational coupling 
parameter $G_{eff}$ \cite{frtHarko,thermo1}.
Then, while the gravitational constant $G$ 
is contemplated as a fundamental constant of nature 
that is considered to be fixed, the gravitational 
coupling parameter $G_{eff}$ is not fixed generally, 
and can vary with time during the evolution of the universe. 
Regarding equation (\ref{eq12}), for the present 
modified $f(R,T)$ model we find $G_{eff}=G(1+\lambda)$,  
and thus, we notice that the effective gravitational 
constant is a function of the $f(R,T)$ model parameter 
$\lambda$. So, we perceive that the expansion history of 
the universe can be influenced by the matter-geometry 
coupling in $f(R,T)$ modified gravity.

On the other hand, modified field equations to linear order of
perturbations are given by
\begin{align}
\frac{a'}{a}h'-2k^2\eta=8\pi G a^2 \Big((1+\lambda)\sum_{i}\delta \rho_{i}-3\lambda\sum_{i}\delta p_{i}\Big) \,, \label{eq13}
\end{align}
\begin{align}
k^2\eta'=4\pi G(1+2\lambda) a^2 \sum_{i}\big(\bar{\rho}_i+\bar{p}_i\big)\theta_{i} \,, \label{eq14}
\end{align}
\begin{align}
\frac{1}{2}h''+3\eta''+\big(h'+6\eta'\big)\frac{a'}{a}-k^2\eta=0 \,, \label{eq15}
\end{align}
\begin{align}
-2\frac{a'}{a}h'-h''+2k^2\eta=24\pi G a^2 \Big(\lambda\sum_{i}\delta \rho_{i}+(1+5\lambda)\sum_{i}\delta p_{i}\Big)  \,, \label{eq16}
\end{align}
where $\theta_i$ in equation (\ref{eq14}) is the 
divergence of velocity perturbations for the component 
\textit{i}th in the universe, and also we neglect the 
anisotropic stress contribution in equation (\ref{eq15}). 
It should be mentioned that choosing $\lambda=0$ recovers
field equations in standard cosmology.

Furthermore, the energy-momentum tensor is not covariantly
conserved in $f(R,T)$ gravity, and then we have
\begin{align} \label{eq17}
\nabla_{\mu}T^{\mu}_{\nu}=-\frac{\lambda}{1+2\lambda}\partial_{\nu}\big(\rho+3p\big) \,.
\end{align}
In this regard, non-conservation equations in $f(R,T)$ model for
the component \textit{i}th of the universe in background and
perturbation levels take the form
\begin{align}
\bar{\rho}'_i+\frac{3(1+w_i)(1+2\lambda)}{1+\lambda(1-3w_i)}\frac{a'}{a}\bar{\rho}_i=0 \,, \label{eq18}
\end{align}
\begin{align}
\delta'_{i}=&\frac{1+2\lambda}{-1+\lambda(-1+3c^2_{si})} \nonumber \\
&\times \Bigg\{\delta_{i}\frac{a'}{a}\bigg(\frac{3(1+w_i)\big(-1+\lambda(-1+3c^2_{si})\big)}{1+\lambda(1-3w_i)} \nonumber \\
&+3(1+c^2_{si})\bigg) \nonumber \\
&+\frac{1}{2}h'(1+w_i)+(1+w_i)\theta_{i} \nonumber \\
&\times \bigg[1+9\frac{(c^2_{si}-c^2_{ai})(1+2\lambda)}{k^2\big(1+\lambda(1-3w_i)\big)} \nonumber \\
&\times \bigg(\Big(\frac{a'}{a}\Big)^2
-\frac{\lambda}{1+2\lambda}\Big(\frac{a''}{a} \nonumber \\
&-\Big(\frac{a'}{a}\Big)^2\Big(1+\frac{3(1+w_i)(1+2\lambda)}{1+\lambda(1-3w_i)}\Big)\Big)\bigg)\bigg] \nonumber \\
&-\frac{9(c^2_{si}-c^2_{ai})(1+w_i)\lambda}{k^2\big(1+\lambda(1-3w_i)\big)}\frac{a'}{a}\theta'_{i} \Bigg\} \,, \label{eq19}
\end{align}
\begin{align}
\theta'_{i}&=\theta_{i}\frac{a'}{a}
\bigg[\frac{3(1+w_i)(1+2\lambda)+3(1+5\lambda)(c^2_{si}-c^2_{ai})}{1+\lambda(1-3w_i)}-4\bigg] \nonumber \\
&+\frac{k^2}{(1+w_i)(1+2\lambda)}\Big(c^2_{si}+\lambda(1+5c^2_{si})\Big)\delta_{i} \,, \label{eq20}
\end{align}
where $c_{si}$ and $c_{ai}$ are the physical sound 
speed and the adiabatic sound speed of the component 
\textit{i}th in the universe, respectively.

Moreover, in pursuance of applying the 
simulated GW data to improve the derived 
constraints on cosmological parameters of 
$f(R,T)$ model from recent observations,
we focus our attention toward the GW
propagation in $f(R,T)$ modified gravity. Accordingly,
without considering the anisotropic stress contribution,
the propagation of tensor perturbations in $f(R,T)$ model
is similar to the GR theory, given by \cite{pgw}
\begin{equation} \label{eq21}
h''_{(+,\times)}+2\frac{a'}{a}h'_{(+,\times)}+k^2h_{(+,\times)}=0 \;,
\end{equation}
with the two plus and cross polarizations. Then, the GW
luminosity distance $d_L^\mathrm{gw}$ would be the 
same as the standard luminosity distance 
$d_L^\mathrm{em}$, written as
\begin{equation} \label{eq22}
d_L^\mathrm{gw}(z)=(1+z)\int_0^z{\frac{\mathrm{d}z}{H(z)}} \,,
\end{equation}
where the Hubble parameter $H(z)$ is expressed in the modified
Friedmann equation (\ref{eq10}), and then we see that the 
GW luminosity distance depends on the $f(R,T)$ model 
parameter $\lambda$.
So, in pursuit of studying the GW luminosity 
distance in $f(R,T)$ model, which is indeed important 
to create the simulated GW data, we should investigate 
the influence of modified $f(R,T)$ gravity on the 
expansion rate $H(z)$ of the universe. Then, we perceive 
that $d_L^\mathrm{gw}$ is determined by the modified 
Friedmann equation (\ref{eq10}) in $f(R,T)$ gravity.
Thus, it is now possible to generate mock GW data from 
LISA SS sources based on the specific $f(R,T)$ gravity 
described in this section as the fiducial model, for the 
sake of constraining the model parameters with observations.
%%%%%%%%%%%%%%%%%%%%%%%%%%%%%%%%%%%%%%%%%%%%%%%%%%%%%%%%%%%%%%%%%%
\section{Results and analysis} \label{sec3}
This section is dedicated to numerical study of the $f(R,T)$ model
described in section \ref{sec2}. For this purpose, we employ an
MCMC\footnote{Markov Chain Monte Carlo} calculation using the
publicly package M\textsc{onte} P\textsc{ython} \cite{mp1,mp2}
to confront model with currently available observations as well
as forecasted GW data.

Considering Ref. \cite{ASH}, we have modified the Boltzmann code
CLASS\footnote{Cosmic Linear Anisotropy Solving System} \cite{cl}
according to the field equations of $f(R,T)$ theory, to study
the evolution of cosmological observables in this model of
modified gravity. As explained in Ref. \cite{ASH}, exploring
matter power spectra diagrams disclosed a structure growth
suppression in $f(R,T)$ model, which proves to be compatible
with local measurements of large scale structures
\cite{s81,s82,s83,s84,s85,s86}. Furthermore, pondering MCMC
results, we noticed that the $\sigma_8$ tension can be relieved
in $f(R,T)$ gravity, while the Hubble tension 
becomes more serious in this theory of modified gravity \cite{ASH}.
Thereupon, in the present work
we exploit three mock catalogs of LISA SS sources 
together with recent observational measurements to 
make improvements on constraining cosmological parameters.

In what follows, we describe the utilized observational probes in
numerical investigation, and further we compare $f(R,T)$ gravity
with observations.
%%%%%%%%%%%%%%%%%%%%%%%%%%%%%%%%%%%%%%%%%%%%%%%%%%%%%%%%%%%%%
\subsection{Observational datasets}
Here, we introduce the cosmological data applied in our MCMC analysis.
Concerning current observations, we employ the Planck 2018 data
including high-$l$ TT,TE,EE, low-$l$ EE, low-$l$ TT, and lensing
measurements \cite{cmb3} (Planck), the Sunyaev-Zeldovich effect
measured by Planck \cite{sz1,sz2} (SZ), the weak lensing data
\cite{lens1,lens2} (WL), the supernovae data from the Pantheon
sample \cite{pan} (SN), the baryon acoustic oscillations data
\cite{bao1,bao2,rsd1,rsd2} (BAO), and also the redshift-space
distortions measurements \cite{rsd1,rsd2} (RSD). Hereafter,
the combined dataset "Planck+SZ+WL+SN+BAO+RSD" is considered
as dataset I.

Moreover, in case of forecast data, 
we generate three LISA SS mock catalogs, assuming the described 
$f(R,T)$ gravity in section \ref{sec2} as the fiducial model. 
LISA is a space-based interferometer which is scheduled to 
detect massive black hole binary (MBHB) coalescences in the 
range $10^3$ to $10^7$ solar masses, 
up to redshift $z\sim 10$ \cite{lisa1,mass}.
Notably, MBHBs are anticipated to produce powerful observable
EM counterparts, since they merge in gas-rich nuclear environments,
and hence are considered as the main SS sources for LISA. So,
the LISA mission provides us with a deep comprehension of galaxy
formation and evolution, as well as fundamental physics.
We follow the method described in Ref. \cite{ssapp3} to generate
mock GW catalog for $f(R,T)$ modified gravity, where the redshift
distribution of SS events is chosen according to \cite{redshd}.
Principally, regarding initial conditions of the massive black hole
population at high redshift, there are two scenarios for MBHB population,
namely light-seed and heavy-seed. The light-seed scenario is
based on the speculation that massive black holes evolve from the
remnants of population III (Pop III) stars, while the heavy-seed
scenario presumes that massive black holes grow from the collapse
of protogalactic disks. 
Moreover, in the heavy-seed scenario, it is possible to 
consider a delay (or no delay) between the galaxy and massive 
black hole mergers, yielding Delay (No Delay) populations.
Thereupon, we can contemplate three
distinct MBHB formation models named as Pop III, Delay, and
No Delay \cite{catalog}. On the other hand, the realistic $1\sigma$
luminosity distance error to MBHB events for LISA is a combination
of weak lensing, peculiar velocity, instrumental, and redshift
uncertainties, taking the form \cite{ssapp3,error}
\begin{equation} \label{eq23}
\sigma_\mathrm{LISA}^2=\sigma_\mathrm{delens}^2+\sigma_\mathrm{v}^2+\sigma_\mathrm{inst}^2+\bigg(\frac{\mathrm{d}}{\mathrm{d}z}(d_L)\sigma_\mathrm{photo}\bigg)^2 \;.
\end{equation}
Concerning the LISA weak lensing error, we have
\begin{equation} \label{eq24}
\sigma_\mathrm{delens}(z)=F_\mathrm{delens}(z)\sigma_\mathrm{lens}(z) \;,
\end{equation}
where
\begin{align}
& F_\mathrm{delens}(z)=1-\frac{0.3}{\pi/2}\arctan\bigg(\frac{z}{0.073}\bigg) \;,  \label{eq25} \\
& \frac{\sigma_\mathrm{lens}(z)}{d_L(z)}=0.066\bigg(\frac{1-(1+z)^{-0.25}}{0.25}\bigg)^{1.8} \;. \label{eq26}
\end{align}
The peculiar velocity error for LISA is given by
\begin{equation} \label{eq27}
\frac{\sigma_\mathrm{v}(z)}{d_L(z)}=\bigg(1+\frac{c(1+z)^2}{H(z)d_L(z)}\bigg)\frac{500\,\mathrm{km/s}}{c} \;.
\end{equation}
The LISA instrumental uncertainty becomes
\begin{equation}
\frac{\sigma_\mathrm{inst}(z)}{d_L(z)}=0.05\bigg(\frac{d_L(z)}{36.6\,\mathrm{Gpc}}\bigg) \;.
\end{equation}
Furthermore, the redshift measurement error take the form
\begin{equation}
\sigma_\mathrm{photo}(z)=0.03(1+z) \;, \quad \text{if} \;\; z>2 \;.
\end{equation}

\begin{figure}
    \centering
    \includegraphics[width=8.5cm]{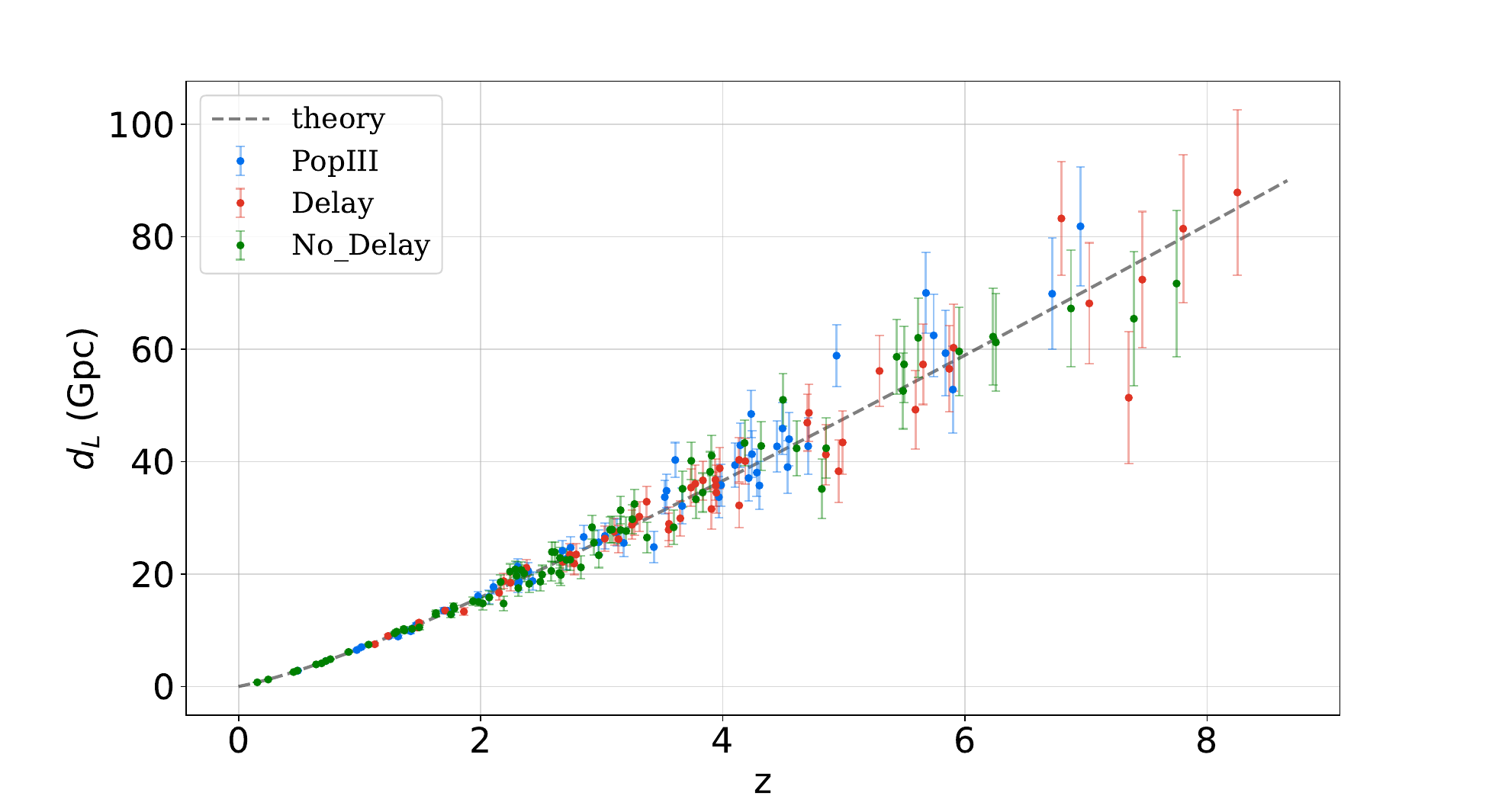}
    \caption{The GW luminosity distance as a function of 
    	redshift, from three generated mock catalogs of LISA SS sources 
    	(light-seed PopIII, heavy-seed Delay, and heavy-seed No Delay)
    	for a ten-year mission. 
    	The "theory" diagram is the f(R,T) model constrained by 
    	dataset I \cite{ASH} considered as the fiducial model.}
    \label{fig1}
\end{figure}

For the purpose of deriving stronger constraints on $f(R,T)$ model
parameters, we aim to generate three SS mock catalogs in accordance
with Pop III, Delay, and No Delay population models, based on a
ten-year LISA mission lifetime. In this respect, the constrained
$f(R,T)$ gravity corresponding to the dataset I studied in Ref.
\cite{ASH} is assumed as the fiducial model, with the best fit
values reported in table \ref{tab1}.
In particular, considering the GW luminosity distance 
described in equation \ref{eq22}, we have used
$H_0=68.43\,\mathrm{km/s/Mpc}$, $\Omega_{\mathrm{M},0}=0.3022$, 
and $\lambda=2.682e-7$ as the fiducial values. 
The created mock catalogs of LISA SS from 
three MBHB populations comparing with the 
theoretical prediction of GW luminosity 
distance are exhibited in figure \ref{fig1}.
Also, It is worth to mention the number of data 
points in three generated mock catalogs of LISA SS sources, 
where Pop III catalog contains 56 GW events, Delay 
population consists of 52 data points, and No Delay 
catalog contains 79 events.

Now that we have presented different exploited datasets in our study,
we proceed to confront the $f(R,T)$ model with current 
cosmological measurements as well as forecasted GW data from 
LISA SS sources.  
%%%%%%%%%%%%%%%%%%%%%%%%%%%%%%%%%%%%%%%%%%%%%%%%%%%%%%%%%%
\subsection{Fit to observations}
This part is devoted to constraints on modified $f(R,T)$ gravity,
applying a combined dataset of current observational measurements
together with mock LISA SS data. To this end, we employ the
cosmological code M\textsc{onte} P\textsc{ython}, where the
corresponding GW likelihoods are included in the code. The
baseline parameter set to be constrained in MCMC analysis
consists of \{$100\,\Omega_{\mathrm{B},0} h^2$,
$\Omega_{\mathrm{DM},0} h^2$, $100\,\theta_s$, $\ln (10^{10}
A_s)$, $n_s$, $\tau_{\mathrm{reio}}$, $\lambda$\}, including
the six $\Lambda$CDM cosmological parameters in addition to
the $f(R,T)$ model parameter $\lambda$. Concerning preliminary
numerical studies, specifically the influence of 
$f(R,T)$ gravity on cosmological observables, namely CMB 
anisotropy and mater power spectra, that is explained in the 
Ref. \cite{ASH}, we choose the prior on $\lambda$ in the range
[$0$, $10^{-4}$]. Moreover, there are four derived parameters
containing the reionization redshift $z_\mathrm{reio}$, the
matter density parameter $\Omega_{\mathrm{M},0}$, the Hubble
constant $H_0$, and the structure growth parameter $\sigma_8$.

\begin{figure}
    \centering
    \includegraphics[width=8.5cm]{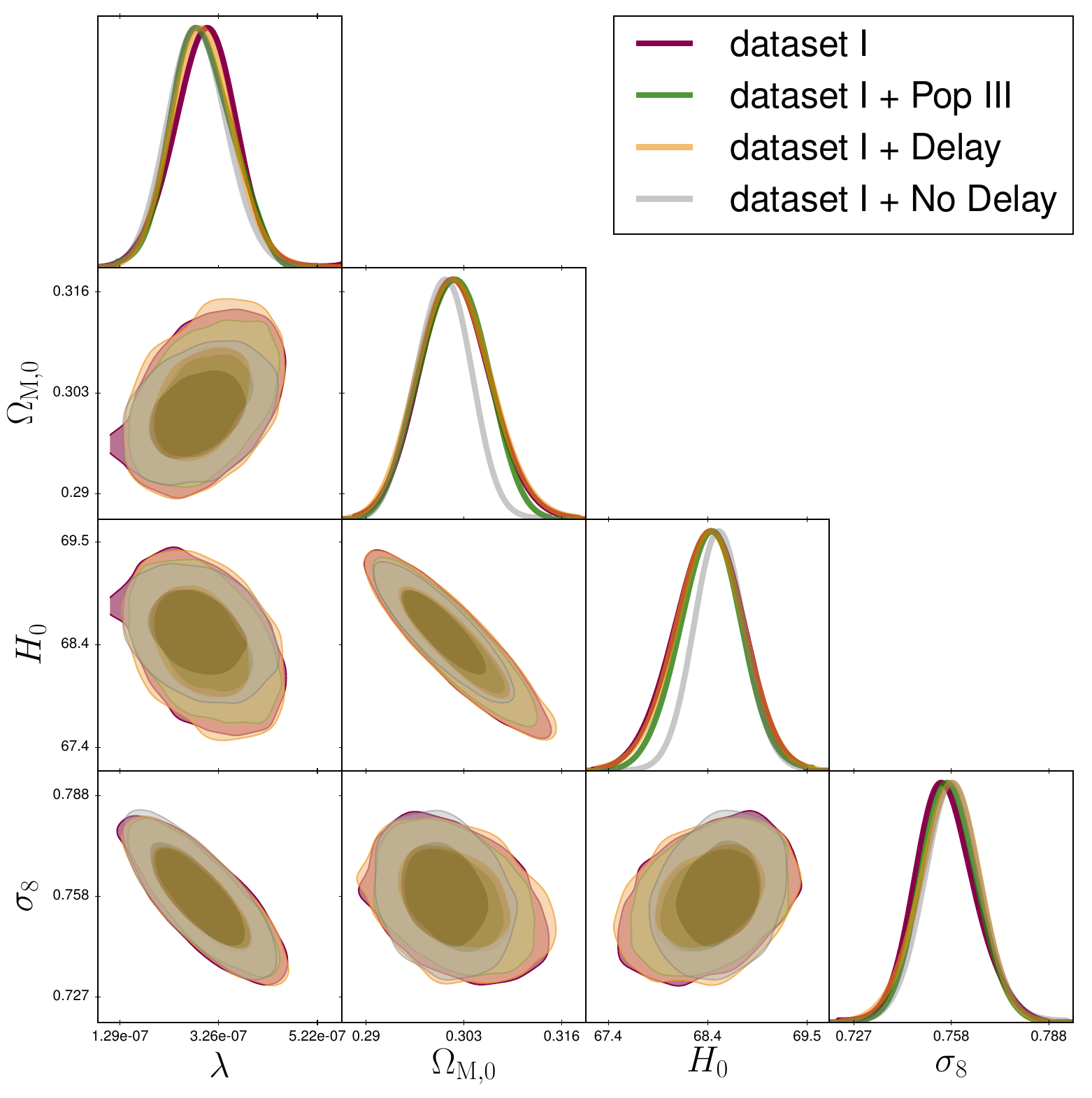}
    \caption{The $1\sigma$ and $2\sigma$ constraints on some selected cosmological parameters of $f(R,T)$ model from "dataset I + Pop III" (green), "dataset I + Delay" (orange), and "dataset I + No Delay" (gray) datasets, compared to dataset I from Ref. \cite{ASH} (purple).}
    \label{fig2}
\end{figure}

In pursuance of improving the obtained constraints on $f(R,T)$
model parameters, we utilize three combined datasets, namely
"dataset I + Pop III", "dataset I + Delay", and
"dataset I + No Delay", where the fitting results are
summarized in table \ref{tab1}.
\begin{table*}
    \centering
    \caption{Observational constraints on $f(R,T)$ gravity from "dataset I + Pop III", "dataset I + Delay", and "dataset I + No Delay" datasets, where constraints from dataset I are also included according to the Ref. \cite{ASH} for comparison.}
    \scalebox{.7}{
        \begin{tabular}{|c|c|c|c|c|c|c|c|c|}
            \hline
            & \multicolumn{2}{|c|}{} & \multicolumn{2}{|c|}{} & \multicolumn{2}{|c|}{} & \multicolumn{2}{|c|}{}\\
            & \multicolumn{2}{|c|}{dataset I} & \multicolumn{2}{|c|}{dataset I + Pop III} & \multicolumn{2}{|c|}{dataset I + Delay} & \multicolumn{2}{|c|}{dataset I + No Delay} \\
            \cline{2-9}
            & & & & & & & & \\
            {parameter} & best fit & 68\% \& 95\% limits & best fit & 68\% \& 95\% limits & best fit & 68\% \& 95\% limits & best fit & 68\% \& 95\% limits \\ \hline
            & & & & & & & & \\
            $100\,\Omega_{\mathrm{B},0} h^2$ & $2.249$ & $2.246^{+0.013+0.027}_{-0.013-0.026}$ & $2.250$ & $2.246^{+0.013+0.025}_{-0.013-0.025}$ & $2.257$ & $2.247^{+0.013+0.025}_{-0.012-0.026}$ & $2.245$ & $2.247^{+0.012+0.024}_{-0.012-0.024}$ \\
            & & & & & & & & \\
            $\Omega_{\mathrm{DM},0} h^2$ & $0.1190$ & $0.1189^{+0.00082+0.0017}_{-0.00081-0.0017}$ & $0.1195$ & $0.1189^{+0.00075+0.0016}_{-0.00078-0.0015}$ & $0.1186$ & $0.1189^{+0.00080+0.0018}_{-0.00086-0.0017}$ & $0.1186$ & $0.1186^{+0.00062+0.0013}_{-0.00071-0.0013}$ \\
            & & & & & & & & \\
            $100\,\theta_s$ & $1.042$ & $1.042^{+0.00027+0.00058}_{-0.00030-0.00060}$ & $1.042$ & $1.042^{+0.00027+0.00056}_{-0.00029-0.00055}$ & $1.042$ & $1.042^{+0.00028+0.00054}_{-0.00028-0.00058}$ & $1.042$ & $1.042^{+0.00027+0.00057}_{-0.00029-0.00055}$ \\
            & & & & & & & & \\
            $\ln (10^{10} A_s)$ & $3.052$ & $3.050^{+0.014+0.032}_{-0.016-0.028}$ & $3.044$ & $3.051^{+0.013+0.028}_{-0.015-0.028}$ & $3.045$ & $3.051^{+0.014+0.029}_{-0.015-0.030}$ & $3.058$ & $3.052^{+0.014+0.029}_{-0.016-0.029}$ \\
            & & & & & & & & \\
            $n_s$ & $0.9664$ & $0.9682^{+0.0035+0.0070}_{-0.0036-0.0068}$ & $0.9682$ & $0.9683^{+0.0036+0.0069}_{-0.0032-0.0070}$ & $0.9703$ & $0.9681^{+0.0035+0.0072}_{-0.0035-0.0069}$ & $0.9660$ & $0.9689^{+0.0032+0.0068}_{-0.0033-0.0067}$ \\
            & & & & & & & & \\
            $\tau_\mathrm{reio}$ & $0.05939$ & $0.05772^{+0.0067+0.015}_{-0.0081-0.015}$ & $0.05573$ & $0.05837^{+0.0064+0.014}_{-0.0074-0.014}$ & $0.05536$ & $0.05802^{+0.0065+0.014}_{-0.0078-0.014}$ & $0.06203$ & $0.05915^{+0.0067+0.015}_{-0.0076-0.014}$  \\
            & & & & & & & & \\
            $\lambda$ & $2.682\mathrm{e}{-7}$ & $2.972\mathrm{e}{-7}^{+6.0\mathrm{e}{-8}+1.3\mathrm{e}{-7}}_{-6.5\mathrm{e}{-8}-1.2\mathrm{e}{-7}}$ & $3.132\mathrm{e}{-7}$ & $2.935\mathrm{e}{-7}^{+5.7\mathrm{e}{-8}+1.2\mathrm{e}{-7}}_{-6.4\mathrm{e}{-8}-1.2\mathrm{e}{-7}}$ & $2.960\mathrm{e}{-7}$ & $2.933\mathrm{e}{-7}^{+5.6\mathrm{e}{-8}+1.3\mathrm{e}{-7}}_{-6.7\mathrm{e}{-8}-1.2\mathrm{e}{-7}}$ & $2.845\mathrm{e}{-7}$ & $2.840\mathrm{e}{-7}^{+5.8\mathrm{e}{-8}+1.2\mathrm{e}{-7}}_{-6.3\mathrm{e}{-8}-1.2\mathrm{e}{-7}}$ \\
            & & & & & & & & \\
            $z_\mathrm{reio}$ & $8.152$ & $7.976^{+0.66+1.4}_{-0.79-1.5}$ & $7.792$ & $8.041^{+0.66+1.4}_{-0.71-1.3}$ & $7.727$ & $8.004^{+0.69+1.4}_{-0.72-1.4}$ & $8.412$ & $8.111^{+0.72+1.5}_{-0.70-1.4}$ \\
            & & & & & & & & \\
            $\Omega_{\mathrm{M},0}$ & $0.3022$ & $0.3021^{+0.0047+0.0095}_{-0.0050-0.0098}$ & $0.3051$ & $0.3017^{+0.0042+0.0087}_{-0.0045-0.0087}$ & $0.2997$ & $0.3019^{+0.0046+0.010}_{-0.0051-0.010}$ & $0.3010$ & $0.3003^{+0.0034+0.0074}_{-0.0039-0.0072}$ \\
            & & & & & & & & \\
            $H_0\;[\mathrm{km\,s^{-1}\,Mpc^{-1}}]$ & $68.43$ & $68.42^{+0.37+0.76}_{-0.37-0.73}$ & $68.22$ & $68.45^{+0.36+0.67}_{-0.33-0.67}$ & $68.64$ & $68.44^{+0.38+0.73}_{-0.36-0.80}$ & $68.46$ & $68.55^{+0.28+0.56}_{-0.28-0.57}$ \\
            & & & & & & & & \\
            $\sigma_8$ & $0.7623$ & $0.7561^{+0.0096+0.021}_{-0.010-0.019}$ & $0.7521$ & $0.7569^{+0.0094+0.018}_{-0.010-0.019}$ & $0.7533$ & $0.7570^{+0.010+0.020}_{-0.0094-0.020}$ & $0.7597$ & $0.7583^{+0.0099+0.019}_{-0.0095-0.020}$ \\
            & & & & & & & & \\
            \hline
        \end{tabular}
    }
    \label{tab1}
\end{table*}
The corresponding two-dimensional contour plots for selected
cosmological parameters of $f(R,T)$ model are also displayed
in figure \ref{fig2}.
According to numerical results, including mock GW data yields 
marginal improvements on constraining cosmological 
parameters of $f(R,T)$ model. Specially, the background parameters, 
mainly $\Omega_{\mathrm{M},0}$ and $H_0$, have been 
slightly better constrained after adding Pop III data to 
recent observations. Further, addition of No Delay data can 
provide marginally better 
constraints on background parameters, while we detect no 
significant improvement in case of Delay data. 

To be more specific, let us ponder the measurement precision of
background parameters, means the $1\sigma$ relative error of
$\Omega_{\mathrm{M},0}$ and $H_0$. The constraint precision
of Hubble constant which is $1.08\%$ according to dataset I, would
slightly improves to $1.01\%$ with introducing the 
Pop III data, and also there is a marginal 
improvement of $0.817\%$ in case of No Delay data.
On the other hand, considering the measurement precision of
$\Omega_{\mathrm{M},0}$ which is $3.21\%$ for dataset I, we
detect a slightly enhanced precision of $2.88\%$ and $2.43\%$ 
after the addition of Pop III and No Delay data, respectively.
Thereupon, we realize that 
forecasted GW data marginally improve
the obtained constraints on cosmological parameters
of $f(R,T)$ gravity.

On the other hand, concerning the model parameter $\lambda$, 
no improvement on observational constraints can be perceived
after the addition of forecasted GW data from three 
mock catalogs of LISA. In order to comprehend this result, 
let us turn our attention to the propagation of GWs in the 
studied $f(R,T)$ model, described in equation (\ref{eq21}). 
Contemplating the GW propagations in $f(R,T)$ gravity, 
we understand that the tensor perturbations (in absence of 
anisotropic stress) in $f(R,T)$ model have the same form as 
in the theory of GR. Then, it means that tensor perturbations 
in absence of anisotropic stress would not be directly affected 
by the $f(R,T)$ model parameter $\lambda$. Moreover, the GW 
luminosity distance explained in equation (\ref{eq22}) is 
the same as the standard luminosity distance, and so the 
impact of modified $f(R,T)$ gravity can be only recognized 
in the expansion rate $H(z)$ defined in the modified Friedmann 
equation (\ref{eq10}). Thus, we perceive that only the background 
Hubble parameter $H$ is influenced by the modified gravity 
model parameter $\lambda$. Accordingly, since the tensor 
perturbations are not directly affected by $\lambda$, one 
can conclude that the forecasted GW data from LISA SS sources 
are not capable to improve observational constraints on the 
$f(R,T)$ model parameter $\lambda$.

Furthermore, from MCMC analysis we perceive that the Hubble 
tension becomes more severe in $f(R,T)$ gravity, which is also 
thoroughly discussed in the Ref. \cite{ASH}.
%%%%%%%%%%%%%%%%%%%%%%%%%%%%%%%%%%%%%%%%%%%%%%%%%%%%%%%%%%%%%%%%%%%
\section{Closing remarks} \label{sec4}
Modified theories of gravity which consider corrections on
gravitational action in GR theory, provides an appropriate and
successful alternative gravitational models to elucidate observed
insufficiencies where the standard model of cosmology is not
capable to explain. In terms of the modifications to GR, one can
assume a non-minimal coupling between matter and curvature,
identified as $f(R,T)$ modified gravity \cite{frtHarko}. The
energy-momentum tensor in this special modification of GR is not
conserved, and consequently, an extra acceleration arises due to
the matter-geometry interaction.

In this paper, we have concentrated on comparing $f(R,T)$ model
with observations, considering the functional form 
$f(R,T)=R+2f(T)$ where $f(T)=8\pi G\lambda T$.
Particularly, we have forecasted the capability of 
mock LISA SS data to improve cosmological constraints on 
$f(R,T)$ model parameters.
We notice that it is important to study the 
impact of modified $f(R,T)$ gravity on the 
expansion history of the universe described 
in the modified friedmann equation (\ref{eq10}), 
in order to create forecasted GW data from LISA SS 
sources based on the fiducial $f(R,T)$ model.
Accordingly, we have regarded the $f(R,T)$
cosmology constrained by current data called dataset I (reported
in table \ref{tab1}) as our fiducial model to generate three
categories of mock SS data, namely Pop III, Delay, and No Delay,
for a ten-year LISA mission. Numerical studies indicate that
utilizing simulated GW data from LISA SS sources 
results in marginally better constraints on cosmological 
parameters of $f(R,T)$ model. 
Notably, obtained constraints 
on the background parameters, mainly matter density parameter 
and the Hubble constant, have 
been slightly improved in case of "dataset I + Pop III", 
and also "dataset I + No Delay". 

On the other hand, we have noticed no significant
improvements on the model parameter $\lambda$ constraints 
after the addition of mock GW data from LISA SS sources
to recent observations. This result can be interpreted 
by considering the propagation of GWs in $f(R,T)$ model 
described in equation (\ref{eq21}), which is similar to 
the one in GR theory. Specifically, tensor perturbations 
in $f(R,T)$ gravity without considering the anisotropic 
stress, are not directly affected by the model parameter 
$\lambda$, and consequently, the GW luminosity distance 
is the same as the standard luminosity distance in GR. 
Also, we notice the direct impact of $f(R,T)$ model 
apparently on the background Hubble parameter $H(z)$ described 
in the modified Friedmann equation (\ref{eq10}), which determines 
the GW luminosity distance (\ref{eq22}). Thus, because the 
$f(R,T)$ model parameter $\lambda$ affects only the background 
level parameter $H$, and would not directly influence the tensor 
perturbations, no considerable improvement on constraining the 
$f(R,T)$ parameter $\lambda$ is expected after introducing the 
simulated GW data from LISA. 

In summary, we have studied the qualification of 
simulated GW data from LISA SS sources to improve observational 
constraints on $f(R,T)$ model parameters. According to numerical 
results, the forecasted GW data from LISA SS sources would 
marginally improve obtained constraints on $f(R,T)$ model 
parameters, mainly $\Omega_{\mathrm{M},0}$ and $H_0$. 
However, we recognize no substantial enhancement on the 
model parameter $\lambda$ constrains after the addition 
of mock GW data from LISA. 

%%%%%%%%%%%%%%%%%%%%%%%%%%%%%%%%%%%%%%%%%%%%%%%%%%%%%%%%%%%%%%%%%%%%
\section*{Declaration of competing interest}
The authors declare that they have no known competing financial interests
or personal relationships that could have appeared to influence the work
reported in this paper.
%%%%%%%%%%%%%%%%%%%%%%%%%%%%%%%%%%%%%%%%%%%%%%%%%%%%%%%%%%%%%%%%%%%%
\section*{Acknowledgments}
We thank Shiraz University Research Council.
This work is based upon research funded by
Iran National Science Foundation (INSF) under project No. 4024184.
%%%%%%%%%%%%%%%%%%%%%%%%%%%%%%%%%%%%%%%%%%%%%%%%%%%%%%%%%%%%%%%%%%%%%
\section*{Data availability}
No new data were generated or analysed in support of this research.
%%%%%%%%%%%%%%%%%%%%%%%%%%%%%%%%%%%%%%%%%%%%%%%%%%%%%%%%%%%%%%%%%%%%
\section*{Code availability}
The modified version of the CLASS code is available under reasonable 
request.
%%%%%%%%%%%%%%%%%%%%%%%%%%%%%%%%%%%%%%%%%%%%%%%%%%%%%%%%%%%%%%%%%%%%

\interlinepenalty=10000
\bibliography{1}

\end{document}